\documentclass[12pt]{article}

\usepackage{epsfig}
\usepackage{latexsym}

\begin{document}

\begin{titlepage}

\baselineskip 24pt

\begin{center}

{\Large {\bf Updates to the Dualized Standard Model on Fermion Masses and 
Mixings}}

\vspace{.5cm}

\baselineskip 14pt

{\large Jos\'e BORDES}\\
jose.m.bordes\,@\,uv.es\\
{\it Departament Fisica Teorica, Universitat de Valencia,\\
  calle Dr. Moliner 50, E-46100 Burjassot (Valencia), Spain}\\
\vspace{.2cm}
{\large CHAN Hong-Mo}\\
H.M.Chan\,@\,rl.ac.uk \\
{\it Rutherford Appleton Laboratory,\\
  Chilton, Didcot, Oxon, OX11 0QX, United Kingdom}\\
\vspace{.2cm}
{\large TSOU Sheung Tsun}\\
tsou\,@\,maths.ox.ac.uk\\
{\it Mathematical Institute, University of Oxford,\\
  24-29 St. Giles', Oxford, OX1 3LB, United Kingdom}

\end{center}

\vspace{.3cm}

\begin{abstract}

The Dualized Standard Model has scored a number of successes in explaining
the fermion mass hierarchy and mixing pattern.  This note contains updates 
to those results including (a) an improved treatment of neutrino oscillation 
free from previous assumptions on neutrino masses, and hence admitting now 
the preferred LMA solution to solar neutrinos, (b) an understanding of the 
limitation of the 1-loop calculation so far performed, thus explaining
the two 
previous discrepancies with data, 
and (c) an analytic derivation and confirmation of 
the numerical results previously obtained. 

\end{abstract}

\end{titlepage}

\clearpage

\baselineskip 14pt

\section{Introduction}

The Dualized Standard Model (DSM) suggested a few years ago \cite{physcons,
dualgen,genmixdsm} has scored some, to us, notable successes in explaining 
the intricate and otherwise mysterious mass and mixing patterns of quarks 
and leptons.  With only 3 adjustable parameters, the model is able to 
reproduce already at 1-loop level the following mass and mixing parameters 
all to within present experimental errors \cite{ckm,phenodsm}: all 9 CKM 
matrix elements $|V_{rs}|$ for quarks \cite{databook}, the 2 MNS lepton 
mixing matrix elements $|U_{\mu 3}|$ and $|U_{e 3}|$ bounded by neutrino 
oscillation experiments with respectively atmospheric \cite{SuperK} and 
reactor neutrinos \cite{Chooz}, plus the 3 mass ratios $m_c/m_t, m_s/m_b$ 
and $m_\mu/m_\tau$ \cite{databook}, together accounting for 8 independent 
parameters of the Standard Model.  However, the treatment so far published,
culminating in \cite{phenodsm} to which we refer the reader for details,
contains some shortcomings which were not apparent at the beginning but 
have since, with time, been recognized and remedied.  In particular, (a) 
some assumptions were made in $\nu$ oscillations which are later found to 
be unnecessary, (b) the limitations of the 1-loop approximation made which
were not recognized before are now seen to explain why the other masses 
and mixing angles: $m_u, m_d, m_e, U_{e2}$, were not well reproduced, while 
(c) the results obtained only numerically before are now seen to follow 
also from simple analytic considerations.  The purpose of this note is 
to remove these old shortcomings, resulting in both a tighter and a more 
transparent scheme, while affording at the same time an analytic check on 
the previous numerical results.  

We begin with a very brief summary of the basic features of the DSM scheme
necessary for our presentation.  For explanations and details, the reader is 
referred to the literature cited.  The DSM scheme is based on a nonabelian 
generalization of electric-magnetic duality \cite{dualsymm} which offers 
an explanation for the existence of exactly 3 fermion generations and a 
suggestion of how the generation symmetry is broken.  This leads in turn 
to the construction of a Higgs potential and Yukawa coupling which give the
tree-level fermion mass matrix a factorized form \cite{physcons} and allows 
its radiative corrections to be calculated.  Further, the fermion mass 
matrix is found to retain a factorized form even after 1-loop radiative 
corrections: 
\begin{equation}
m = m_T \left( \begin{array}{c} x \\ y \\ z \end{array} \right) (x, y, z),
\label{massmat}
\end{equation}   
where only the normalization $m_T$ depends on the fermion species, 
i.\ e.\ 
$U$ or $D$ type quarks, or charged leptons $L$ or neutrinos $N$.  
The vector $(x, y, z)$, however, now rotates in generation space with 
changing scales $\mu$, and satisfies the RG equation:
\begin{equation}
\frac{d}{d(\ln \mu^2)} \left( \begin{array}{c} x \\ y \\ z 
   \end{array} \right)
   =  \frac{3}{64 \pi^2} \rho^2 \left( \begin{array}{c}
         \tilde{x}_1 \\ \tilde{y}_1 \\ \tilde{z}_1 \end{array} \right),
\label{runxyz}
\end{equation}
with
\begin{equation}
\tilde{x}_1 = \frac{x(x^2-y^2)}{x^2+y^2} + \frac{x(x^2-z^2)}
   {x^2+z^2}, \ \ \ {\rm cyclic},
\label{x1tilde}
\end{equation}
and $\rho^2$ the Yukawa coupling strength \cite{ckm}\footnote{There was
an error in the coefficient of the right-hand side of eq. (\ref{runxyz})
as given in \cite{ckm}, which means that the numerical value of $\rho$ 
given in \cite{ckm} and \cite{phenodsm} should be multiplied by a factor
of $\sqrt{5/3}$.  Other results in these references are not affected.}.  
In (\ref{runxyz}), one sees that the vector $(x, y, z)$, taken to be
normalized from now on, 
is stationary in
orientation with respect to change in $\mu$ at $(x, y, z) = (1, 0, 0)$ and 
$\frac{1}{\sqrt{3}}(1, 1, 1)$, which means that these are to be interpreted 
as fixed points of rotation.  Further, the sign of the derivative is such 
that as $\mu$ changes, the vector ${\bf r} = (x, y, z)$ traces out a 
trajectory on the unit sphere in 3-D generation space starting from the 
``high-energy'' fixed point $(1, 0, 0)$ at infinite scale and ending at 
the ``low-energy'' fixed point $\frac{1}{\sqrt{3}}(1, 1, 1)$ at zero scale.

The fact that the mass matrix rotates with changing scales means that the
usual definition of the 3 generation states as mass eigenstates has to be
refined since it will have to be specified at what scales they are to be
so defined.  An analysis of the situation, defining each state at its own
mass scale, leads to the following conclusion \cite{cevidsm}.  For each 
fermion species $T$ (i.e. whether $U, D, L$ or $N$) the state vectors for 
the 3 generations form an orthonormal triad in generation space which is 
given in terms of the rotating vector ${\bf r}(\mu)$ as follows:
\begin{eqnarray}
{\bf v}_1 & = & {\bf r}_1, \nonumber \\
{\bf v}_2 & = & -\frac{{\bf r}_1 \wedge ({\bf r}_1 \wedge {\bf r}_2)}
   {|{\bf r}_1 \wedge ({\bf r}_1 \wedge {\bf r}_2)|}, \nonumber \\
{\bf v}_3 & = & \frac{{\bf r}_1 \wedge {\bf r}_2}{|{\bf r}_1 \wedge {\bf r}_2|}
\label{vtriad}
\end{eqnarray}
where ${\bf r}_i = {\bf r}(m_i)$ is the rotating vector ${\bf r}(\mu)$ taken
at the scale $\mu = m_i$, with $m_i$ being the physical mass of the $i$th
generation labelled from the heaviest (1) to the lightest (3).  The elements 
of the mixing matrix (whether CKM \cite{CKM} for quarks or MNS \cite{MNS}
for leptons) are then given as the scalar products between the up-states 
${\bf v}_i$ and the down states ${\bf v}'_j$, thus:
\begin{equation}
V_{ij} = {\bf v}_i.{\bf v}'_j.
\label{mixingmat}
\end{equation}
Furthermore, for the fermion species $U, D$ and $L$, the mass of the $i$th 
generation state is given as the solution for $m_i$ to the equation:
\begin{equation}
m_i = m_T |{\bf r}_i.{\bf v}_i|^2.
\label{masses}
\end{equation}
This criterion for determining the masses of the 3 generations, however,
does not apply to neutrinos which, because of a likely see-saw mechanism 
\cite{seesaw}, may have physical masses different from the Dirac masses 
appearing in the above equation and hence require a special treatment to 
be explained later.  With the above formulae, once given the trajectory
for ${\bf r}$, mixing matrix elements and mass ratios between generations
can be evaluated, excepting for the moment where neutrinos are involved.  

\begin{figure}
\hspace*{-1.5cm}
\includegraphics[angle=-90, width=1.2\textwidth]{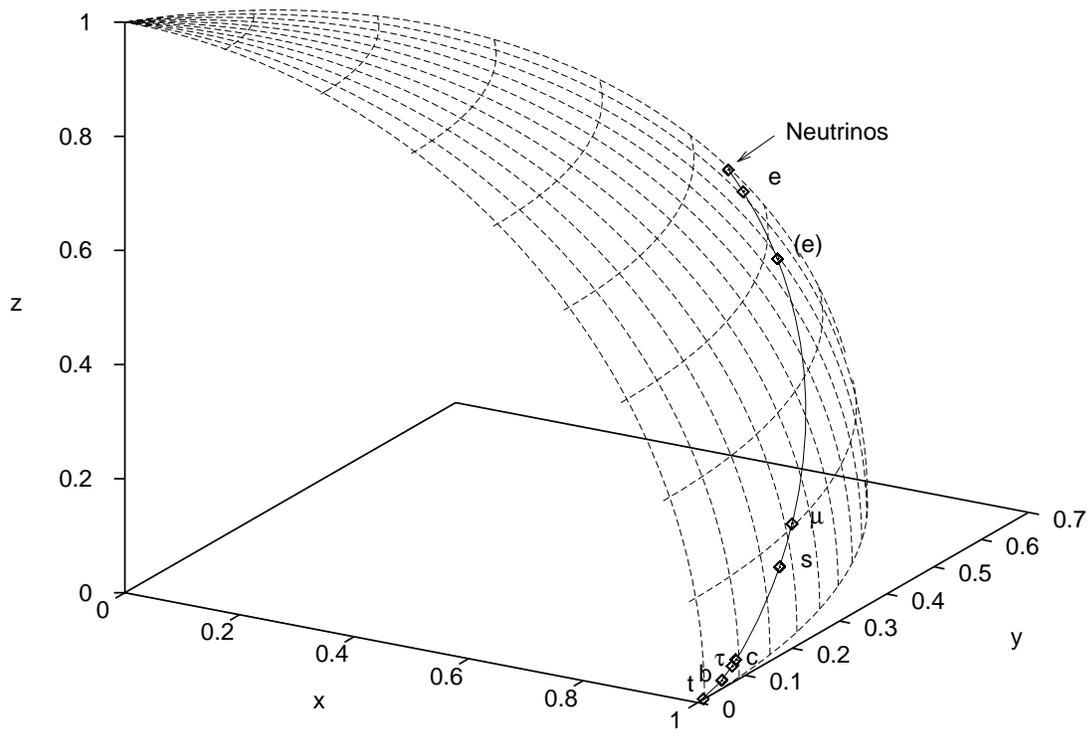}
\caption{Trajectory traced out by the vector ${\bf r}(\mu)$ on the unit
sphere as calculated in \cite{phenodsm} with the various fermion states
marked each at its own mass scale.}
\label{florosphere}
\end{figure}

The trajectory for ${\bf r}(\mu)$ depends on only 3 parameters, two of 
them corresponding to the vacuum expectation values of the Higgs fields 
which specify the rotation trajectory, and the third being the Yakawa 
coupling strength $\rho$ in (\ref{runxyz}) which governs the rotation 
speed.  The first 2 parameters are independent of the fermion species 
$T$, while the third $\rho$ can in principle depend on $T$ but, for 
consistency with the above prescription for defining masses of the 
generations, it was found that $\rho$ has also to be $T$-independent 
(see \cite{ckm} and later).  In \cite{phenodsm}, these 3 parameters were 
fitted to $m_c/m_t, m_\mu/m_\tau$ and the Cabibbo angle, which then 
allows one to calculate the whole trajectory via (\ref{runxyz}).  The 
result is shown in Figure \ref{florosphere} where from the location of 
the various states each marked at its own mass scale, the rotation speed 
with respect to $\mu$ can be gauged.  From this, the CKM matrix for quarks 
was calculated using (\ref{mixingmat}), and gave, as already mentioned, 
all elements within present experimental limits.

\section{$\nu$ Oscillations}

Neutrinos require a special treatment because of the see-saw mechanism 
which is likely to give them physical masses different from the Dirac
mass appearing in the mass matrix (\ref{massmat}).  On the one hand, 
the state vectors for neutrinos, in parallel to other fermions, ought
to be defined at the scales of their physical masses, while on the other
the equation (\ref{masses}) applies only to states whose physical and Dirac 
masses coincide, and hence not directly to neutrinos.  It would appear
therefore that some additional assumption will have to be made on the
relationship between the physical and Dirac masses of neutrinos before
the above scheme can be applied, and this was the approach we took in
\cite{nuosc}.  

It turns out, however, that as far as reproducing the MNS mixing matrix 
elements is concerned, which is at present our main conern, then no 
such new assumption is needed.  This arises as follows.  The electron 
neutrino mass is now restricted by e.g. tritium decay experiments to 
below 3 eV \cite{databook}, while oscillation experiments on solar 
\cite{SuperK,SNO} and atmospheric \cite{SuperK} neutrinos limit the 
mass differences between generations to less than 0.05 eV, implying
that all 3 generations have masses of at most order eV.  According
to the calculation cited in Figure \ref{florosphere}, however, at scales
of order eV, the vector ${\bf r}$ is already so close to the low energy
fixed point $\frac{1}{\sqrt{3}}(1, 1, 1)$ as to be indistinguishable in
the figure.  Referring next to the definition of the state vectors in
(\ref{vtriad}), one sees that in this situation when all 3 generations 
are close together, the 3 vectors in the triad become respectively just 
the position vector, the tangent vector to the trajectory, and the vector 
normal to both the above, all taken at the same point.  In other words, 
one has in particular for the state vector of the heaviest neutrino 
$\nu_3$ just:
\begin{equation}
{\bf v}_1 = {\bf v}_{\nu_3} = {\bf r}_0 = \frac{1}{\sqrt{3}}(1, 1, 1).
\label{vnu3}
\end{equation}
(Notice that neutrinos are conventionally labelled in the opposite order
to that adopted in (\ref{vtriad}) above, i.e. from the lightest ($\nu_1$) 
to the heaviest ($\nu_3$)).

Charged leptons being ordinary Dirac particles, their state vectors can
be calculated by the method detailed in the preamble above in the same 
manner as for quarks.  The result of the calculation performed in 
\cite{phenodsm} and cited in Figure \ref{florosphere} gave:
\begin{eqnarray}
|\tau \rangle & = & (0.996732, 0.076223, 0.026756), \nonumber\\
|\mu \rangle  & = & (-0.075925, 0.774100, 0.628494), \nonumber\\
|e \rangle    & = & (0.027068, -0.628482, 0.777354).
\label{leptonvs}
\end{eqnarray}
From these and (\ref{vnu3}) above, using (\ref{mixingmat}), one easily
obtains the MNS mixing matrix elements:
\begin{eqnarray}
U_{\mu 3} = \langle \mu|\nu_3 \rangle & = & 0.7660, \nonumber\\ 
U_{e 3}   = \langle e  |\nu_3 \rangle & = & 0.1016,
\label{Umu3e3th}
\end{eqnarray}
which are seen to be within the experimental limits obtained for these
quantities by respectively the atmospheric \cite{SuperK} and reactor
\cite{Chooz} neutrino experiments:
\begin{eqnarray}
|U_{\mu 3}| & \sim & 0.56 - 0.83, \nonumber\\ 
|U_{e 3}|   & \sim & 0.00 - 0.15.
\label{Umu3e3ex}
\end{eqnarray}

One notices that the above result is independent of any assumptions 
about the Dirac masses of neutrinos, in contrast to the previous 
treatment of \cite{nuosc}.  In particular, because of the assumptions
made then on neutrino masses, the treatment in \cite{nuosc} is valid
only for the range of masses permitted by the ``vacuum oscillation'' 
solution to the solar neutrino problem.  The present result in
(\ref{Umu3e3th}), on the other hand, having been derived independently 
of assumptions on neutrino masses, is valid for any solution to the
solar neutrino problem, in particular for the large mixing angle MSW 
\cite{MSW} solution.  This is of great practical significance, given 
the increasing preference of recent data for the LMA solution to the
exclusion of the others \cite{LMA,kamland}.

In principle, one can use the trajectory calculated in \cite{phenodsm}
and displayed in Figure \ref{florosphere} to evaluate also the tangent 
and normal vectors at the low energy fixed point ${\bf r}_0 = \frac{1}
{\sqrt{3}}(1, 1, 1)$, namely the state vectors ${\bf v}_{\nu_2}$ and
${\bf v}_{\nu_1}$, and hence all the other elements of the MNS mixing
matrix, in particular the solar neutrino angle $U_{e2}$.  However, as
we shall see in the next section, due to the limitations of the 1-loop
calculation so far performed, such results would be unreliable.

\section{Limitations of 1-loop}

The 1-loop approximation with the parameters fitted as above and as in
\cite{phenodsm} is valid only near the fixed point $(1, 0, 0)$ where 
the rotation is slow.  Away from the fixed point, the RGE will receive
large logaritmic terms from higher-loop contributions.  As seen in Figure 
\ref{florosphere}, however, down to the top mass scale at 175 GeV, the 
vector ${\bf r}$ has rotated from the fixed point at infinite scale by 
an angle of only about 0.012 radians, while even from the top mass scale 
to that at the muon mass at 105 MeV, the vector has rotated by only about 
another 0.3 radians.  For scales above the muon mass, therefore, one can 
hope a 1-loop approximation to have some validity.  But at scales much 
below the $\mu$ mass, the 2-loop correction, which is expected to be of 
the order of the square of the 1-loop result, will become increasingly 
significant, thus making the 1-loop result unreliable.   

Given then the limited validity range for the 1-loop calculation so 
far performed, one has to re-examine the results obtained to ascertain
which could be regarded as reliable.  Let us accept for present
purposes the 1-loop approximation as sufficiently accurate for scales
down to about the muon mass, in which range would then be included 
the 2 heavier generations of the 3 fermion species $U, D, L$, 
i.e. all species except the neutrinos.  We notice from (\ref{vtriad}), 
however, that for $U, D, L$, the triad of state vectors for all 3 
generations are in fact already determined in this range, the vector 
${\bf v}_3$ for the lightest generation being given as the normal 
once the vectors for the 2 heavier generations are specified.  Hence,
given the triads for both the $U$ and $D$ quarks, one can evaluate
the whole CKM matrix with confidence.  Indeed, this is born out by
the comparison given in \cite{phenodsm} of the calculation result:  
\begin{eqnarray}
\lefteqn{
\left( \begin{array}{ccc} |V_{ud}| & |V_{us}| & |V_{ub}| \\
                          |V_{cd}| & |V_{cs}| & |V_{cb}| \\
                          |V_{td}| & |V_{ts}| & |V_{tb}|  
   \end{array} \right)
   =} \nonumber\\
&& \left( \begin{array}{lll} 
   0.9745-0.9762 & 0.217-0.224 & 0.0043-0.0046 \\
   0.217-0.224 & 0.9733-0.9756 & 0.0354-0.0508 \\
   0.0120-0.0157 & 0.0336-0.0486 & 0.9988-0.9994  \end{array} \right),
\label{thckm}
\end{eqnarray}
with the then available experimental numbers quoted from the databook 
\cite{databook}:
\begin{eqnarray}
 \left( \begin{array}{lll} 
   0.9745-0.9760 & 0.217-0.224 & 0.0018-0.0045 \\
   0.217-0.224 & 0.9737-0.9753 & 0.036-0.042 \\
   0.004-0.013 & 0.035-0.042 & 0.9991-0.9994  \end{array} \right),
\label{exckm}
\end{eqnarray} 
which are seen also to be within the most recent limits \cite{data03}.
On the other hand, the masses for the lightest generation, according
to (\ref{masses}), requires running the vector ${\bf r}$ down to the
mass scales of the lightest generation, and hence cannot be achieved
reliably by the 1-loop approximation.  And indeed, the value obtained
in \cite{phenodsm} for the electron mass was 6 MeV, to be compared 
with the correct empirical value of 0.5 MeV.  The mass values for the
light quarks $u, d$ were also not well reproduced, but that may be
due to intrinsic ambiguities in defining light quark masses.

Next, turning to the MNS mixing matrix for leptons, we note first 
that the state vectors for all 3 charged leptons $\tau, \mu, e$ as 
quoted in (\ref{leptonvs}) ought, according to the above criterion,
be reliable.  The same, however, cannot be said for the state vectors
of neutrinos, the mass scales of which are all way outside the range
of validity of the 1-loop calculation.  The only exception is the
vector for the heaviest neutrino $\nu_3$ which, as noted in the 
previous section, can be identified to a good approximation with the
low energy fixed point ${\bf r}_0 = \frac{1}{\sqrt{3}}(1, 1, 1)$ known
to be valid beyond 1-loop \cite{2loop}.  In consequence, one obtains
the result (\ref{Umu3e3th}) in good agreement with experiment.  The 
other 2 vectors in the neutrino triad both depend on the tangent
vector to the trajectory and hence cannot be reproduced by the 1-loop
calculation.  Indeed, the value obtained in \cite{phenodsm} for the 
solar neutrino angle $U_{e2}$, which depends on the state vector for 
the second heaviest neutrino $\nu_2$, is about 0.23 and lies outside
the experimental limits of about 0.4 to 0.7.

In other words, the results of the 1-loop calculation on fermion mass 
and mixing parameters have now all been checked to agree with present 
experiment in the range they are expected to be valid, and are seen 
to deviate from data only outside the expected validity range. 

It should be emphasized, however, that this apparent success achieved 
by calculating only the diagram with a single (dual colour) Higgs loop 
makes sense only if one regards both the mass hierarchy and mixing as 
consequences of the mass matrix rotation.  As far as the orientation of
the rotating vector ${\bf r}(\mu)$ rotation is concerned, it was shown 
in \cite{ckm} that radiative corrections due to standard model particles 
are zero, and those due to other dual colour particles are small, leaving 
thus the calculated (dual colour) Higgs loop with no competition.  The 
same approximation, however, would not be applicable, for example, to the 
normalization of the rotating vector.   

\section{Analytic Approach}

Previous calculations with the DSM scheme, e.g. \cite{phenodsm}, were done 
numerically, on which calculations the above remarks also rely.  It is 
found, however, that with some simple yet quite reasonable approximations, 
an analytic approach is possible which provides first a check on the 
numerical results, and second an easier means for the reader to scrutinize 
and verify them.  It also gives algebraic relations between fermion mixing 
elements and mass ratios which may be useful for certain purposes.

First, let us parametrize the vector ${\bf r}(\mu)$ in terms of the usual
polar co-ordinates, thus:
\begin{equation}
{\bf r}(\mu) = (\cos \theta(\mu), \sin \theta(\mu) \cos \phi(\mu),
   \sin \theta(\mu) \sin \phi(\mu)),
\label{polarco}
\end{equation}
and denote by $\theta_i, \phi_i$ the corresponding angles taken at the
scale $\mu = m_i$ with $m_i$ being the mass of the state $i$.  Next, we 
recall from the last section that the 1-loop approximation is valid only 
in the region near the high energy fixed point $(1, 0, 0)$ so that there
is no loss of generality at 1-loop to make a small angle approximation,
i.e. assuming that all $\theta$'s are small and that, although the $\phi$'s
are in principle arbitrary, their differences are small.  In the same 
spirit, we assume also that $\theta_1$ is small compared to $\theta_2$.  
In this approximation then, one has:
\begin{eqnarray}
{\bf v}_1 & = & (1 - \frac{\theta_1^2}{2}, \theta_1 \cos \phi_1,
   \theta_1 \sin \phi_1), \nonumber \\
{\bf v}_2 & = & (0, \cos \phi_2, \sin \phi_2) \nonumber \\
          &   & + \theta_1 (-\cos(\phi_2 - \phi_1), -\frac{1}{\theta_2}
                \sin(\phi_2 - \phi_1) \sin \phi_2, \frac{1}{\theta_2}
                \sin(\phi_2 - \phi_1) \cos \phi_2), \nonumber \\
{\bf v}_3 & = & (0, -\sin \phi_2, \cos \phi_2) \nonumber \\
          &   & + \theta_1 \sin(\phi_2 - \phi_1) (1, -\frac{1}{\theta_2}
                \cos \phi_2, -\frac{1}{\theta_2} \sin \phi_2).
\label{vapprox}
\end{eqnarray}

Using (\ref{vapprox}) and (\ref{mixingmat}), one obtains the approximate 
forms of the CKM matrix elements:
\begin{eqnarray}
V_{tb} & = & 1 - \frac{1}{2} (\theta_b - \theta_t)^2, \nonumber \\
V_{cs} & = & \cos(\phi_s - \phi_c) + \sin(\phi_s - \phi_c)
             \left\{\frac{\theta_t}{\theta_c} \sin(\phi_c - \phi_t)
             - \frac{\theta_b}{\theta_s} \sin(\phi_s - \phi_b) \right\},
             \nonumber \\ 
V_{ud} & = & - V_{cs}, \nonumber \\
V_{cd} & = & \sin(\phi_s - \phi_c) + \cos(\phi_s - \phi_c)
             \left\{\frac{\theta_b}{\theta_s} \sin(\phi_s - \phi_b)
             - \frac{\theta_t}{\theta_c} \sin(\phi_c - \phi_t) \right\}, 
             \nonumber \\
V_{us} & = & - V_{cd}, \nonumber \\
V_{ts} & = & -\theta_b \cos(\phi_s - \phi_b)+\theta_t \cos(\phi_s -\phi_t),
             \nonumber \\
V_{cb} & = & \theta_b \cos(\phi_c - \phi_b)-\theta_t \cos(\phi_c - \phi_t),
             \nonumber \\
V_{td} & = & \theta_b \sin(\phi_s - \phi_b)-\theta_t \sin(\phi_s - \phi_t),
             \nonumber \\
V_{ub} & = & -\theta_b \sin(\phi_c - \phi_b)+\theta_t \sin(\phi_c -\phi_t).
\label{Vapprox}
\end{eqnarray}
From these formulae, one deduces immediately that:
\begin{eqnarray}
|V_{tb}| & \sim & \cos \Delta\theta \sim 1, \nonumber \\
|V_{cs}| & = & |V_{ud}| \sim \cos (\Delta \phi) \sim 1, \nonumber \\
|V_{cd}| & = & |V_{us}| \sim - \sin (\Delta \phi) \sim {\rm small}, 
\nonumber \\
|V_{ts}| & = & |V_{cb}| \sim \Delta \theta \sim {\rm small}, \nonumber \\
|V_{td}| & = & |V_{ub}| \sim \theta \sin(\Delta \phi) \sim {\rm very\ small}.
\label{Vorder}
\end{eqnarray}
In other words, simply on the premises of a rotating mass matrix as given
in (\ref{massmat}) without using even the evolution equation (\ref{runxyz}) 
for the rotating vector ${\bf r}$ apart from the condition that ${\bf r}$ 
should remain near the high energy fixed point $(1, 0, 0)$, one has already 
derived the well-known, and long-wondered at, hierarchy of CKM elements as
observed in experiment.  This result was anticipated already using some 
elementary differential geometry in \cite{features} but is here now made 
completely explicit.

Next, let us rewrite the evolution equation (\ref{runxyz}) for the 
unnormalized vector $(x,y,z)$ as:
\begin{equation}
 \frac{1}{x} \frac{dx}{dt} =f_1,\quad \frac{1}{y} \frac{dy}{dt} =f_2, 
\quad \frac{1}{z} \frac{dz}{dt} =f_3,
\end{equation}
with $t=\ln \mu,\ k=3 \rho^2/(32 \pi^2)$, where
\begin{equation}
f_1=k \left(\frac{x^2-y^2}{x^2+y^2} + \frac{x^2-z^2}{x^2+z^2}\right),\ {\rm
cyclic}.
\end{equation}
To get rid of the normalization, we take differences of these
equations, so that we can still use the parametrization
(\ref{polarco}), which, with arbitrary $\phi$ and small $\theta$, becomes:
\begin{equation}
{\bf{r}} \simeq (1, \theta \cos \phi, \theta \sin \phi),
\end{equation}
and
\begin{equation}
f_1 \simeq 2k,\ f_2 \simeq k(\cos 2 \phi -1), \ f_3 \simeq -k(1+\cos 2 \phi).
\end{equation}
Hence we get for $f_1-f_3$ and $f_2-f_3$ respectively:
\begin{eqnarray}
-\frac{1}{\theta \sin \phi} \frac{d(\theta \sin \phi)}{dt} & = & k(3 +
\cos 2 \phi), \\
\frac{1}{\theta \cos \phi} \frac{d(\theta \cos \phi)}{dt} -
\frac{1}{\theta \sin \phi} \frac{d(\theta \sin \phi)}{dt} & = & 2k
\cos 2 \phi, 
\end{eqnarray}
which simplify to
\begin{eqnarray}
\frac{1}{\sin 2 \phi \cos 2 \phi} \frac{d \phi}{dt} & = & -k,\\
\frac{1}{\theta} \frac{d \theta}{dt} & = & -k(2+\sin^2 2 \phi),
\end{eqnarray}
giving
\begin{eqnarray}
\frac{\tan 2 \phi}{\tan 2 \phi_0} & = &
\left(\frac{\mu_0}{\mu}\right)^{2k}, \label{phieqn} \\
\frac{\theta}{\theta_0} & = & \left(\frac{\mu_0}{\mu}\right)^{2k}
\left( \frac{\cos 2\phi_0}{\cos 2 \phi}\right)^{1/2}. \label{thetaeqn}
\end{eqnarray}
Equation (\ref{thetaeqn}) reduces to
\begin{equation}
\frac{\theta}{\theta_0} =  \left(\frac{\mu_0}{\mu}\right)^{2k}
\label{thetaappr} 
\end{equation}
if we further assume that $\phi^2$ is small.

With this solution, one easily checks that, for consistency with 
(\ref{masses}), $\rho$ has to be the same for all fermion species $T$ to 
a good approximation, a result first found in \cite{ckm} numerically, as 
follows.  From (\ref{masses}), one has:
\begin{equation}
m_2 = m_1 \sin^2 (\theta_2 - \theta_1),
\label{m2fromm1}
\end{equation}
or else, recalling  $\theta_2 \gg \theta_1$, 
which is an approximation numerically 
similar to that for obtaining the linearized solution (\ref{thetaappr})
above:
\begin{equation}
m_2 \sim m_1 \theta_2^2.
\label{m2fromm1a}
\end{equation} 
Hence, using (\ref{thetaappr}), one has:
\begin{equation}
\sqrt{\frac{m_2}{m_1}} \sim \theta_I \left( \frac{\mu_I}{m_2} \right)^{2k},
\label{m2overm1}
\end{equation}
with $\theta_I$ being the value of $\theta$ at some chosen (large) initial 
value $\mu_I$.  Applying the formula (\ref{m2overm1}) to successively the 
fermion species $U, D, L$ and eliminating $\theta_I$, one obtains:
\begin{eqnarray}
\left( \frac{m_c m_b}{m_t m_s} \right)^{1/2} & = & 
  \left(\frac{\mu_I}{m_c}\right)^{2k_U} \left(\frac{m_s}{\mu_I}\right)^{2k_D}, 
  \nonumber \\
\left( \frac{m_c m_\tau}{m_t m_\mu} \right)^{1/2} & = & 
\left(\frac{\mu_I}{m_c}\right)^{2k_U} \left(\frac{m_\mu}{\mu_I}\right)^{2k_L},
\label{mconsist}
\end{eqnarray}
which is consistent for arbitrary $\mu_I$ only when $k_U = k_D = k_L$, as
required.  We note that this result is independent of the choice of the 3 
model parameters.

Turning next to parameter-dependent results, we shall first determine the
parameter values by fitting them to the 3 best known empirical quantities
as was done in \cite{phenodsm}, namely the 2 mass ratios $m_c/m_t$ and
$m_\mu/m_\tau$, and the Cabbibo angle.  From (\ref{m2overm1}) using
the $t$ scale as $I$, 
one deduces 
the approximate relations:
\begin{eqnarray}
\frac{m_c}{m_t} & = & \theta _t^2 \left( \frac{m_t}{m_c} \right)^{4k}, 
   \nonumber \\
\frac{m_\mu}{m_\tau} & = & \theta_t^2 \left( \frac{m_t}{m_\mu} \right)^{4k}.
\label{mcmmu}
\end{eqnarray}
Inputting then the empirical values in GeV of $m_t \sim 175, m_c \sim 1.25,
m_\tau \sim 1.777, m_\mu \sim 0.105$, one easily obtains;
\begin{equation}
k \sim 0.21, \ \ \ \theta_t \sim 0.011.
\label{ktheta}
\end{equation}
These values compare very well with the values $k \sim 0.20, \theta_t \sim
0.012$ obtained before from the numerical fit of \cite{phenodsm}, affording
thus a check on both the present approximation and the previous numerical
result.  The good agreement is actually a little fortuitous, 
being beyond what can be expected from the crudeness of the approximations 
made in deriving the formulae used above.

With now $k$ and $\theta_t$ as 2 of our parameters, we can next proceed to 
evaluate other mass ratios and mixing matrix elements which depend only on
$\theta$.  For mass ratios within the range of validity, we have only
one relation (from (\ref{mconsist}) using $m_b=4.2$ GeV):
\begin{equation}
m_s \sim m_c \left( \frac{m_b}{m_t} \right)^{1/(4k + 1)},
\label{ms}
\end{equation}
giving $m_s \sim 160 \ {\rm MeV}$ (cf. 75 to 170 MeV \cite{databook}).  For
the CKM matrix elements, we have:
\begin{equation}
V_{tb} \sim 1 -\theta_b^2/ 2 \sim 1 - \frac{1}{2} \frac{m_c}{m_t}
   \left(\frac{m_c}{m_b}\right)^{4k},
\label{vtba}
\end{equation}
giving $V_{tb} \sim 1 - 0.001$ (cf. 0.9990 to 0.9993 \cite{databook}), and
\begin{equation}
V_{ts} = - V_{cb} \sim -\theta_b + \theta_t \sim - \sqrt{\frac{m_c}{m_t}}
   \left(\frac{m_c}{m_b}\right)^{2k},
\label{vtsa}
\end{equation}
giving $V_{ts} = - V_{cb} \sim 0.05$ (cf. 0.035 to 0.043 \cite{databook}).
These number also agree well with those obtained from the numerical result
of \cite{phenodsm} quoted in (\ref{thckm}).

Inputting next the Cabbibo angle $V_{us} \sim 0.22$ to determine the last
parameter which we can take as $\sin (\phi_s - \phi_c)$ and which for small
$\phi$ differences, according to (\ref{Vapprox}), can be take approximately 
as:
\begin{equation}
\sin (\phi_s - \phi_c) = \frac{V_{us}}{1 + \left( m_s/m_b \right)^{2k}},
\label{phisc}
\end{equation}
we obtain $\sin (\phi_s - \phi_c) \sim 0.18$.  With this value fixed, we 
can now evaluate the remaining CKM matrix elements, thus:
\begin{equation}
V_{ud} \sim - V_{cs} \sim 1 - \frac{1}{2} V_{us}^2,
\label{vuda}
\end{equation}
giving $V_{ud} \sim - V_{cs} \sim 0.976$ (cf. 0.9734 to 0.9749 
\cite{databook}),
\begin{equation}
V_{td} \sim \theta_b \sin (\phi_s - \phi_c) \sim \sqrt{\frac{m_c}{m_t}}
   \left( \frac{m_c}{m_b} \right)^{2k} \frac{V_{us}}
   {1 + \left (m_s/m_b \right)^{2k}},
\label{vtda}
\end{equation}
giving $V_{td} \sim .009$ (cf. 0.004 to 0.014 \cite{databook}), and
\begin{equation}
V_{ub} \sim \theta_b \frac{\theta_c-\theta_b}{\theta_s} \sin(\phi_s-\phi_c)
   \sim \sqrt{\frac{m_c}{m_t}} \left( \frac{m_s}{m_b} \right)^{2k}
   \frac{V_{us}}{1 + \left( m_s/m_b \right)^{2k}},
\label{vuba}
\end{equation}
giving $V_{ub} \sim 0.004$ (cf. 0.002 to 0.005 \cite{databook}).  Again,
these estimates agree well with the values (\ref{thckm}) obtained before
numerically in \cite{phenodsm}.  Contrary to usual belief the CKM
matrix elements appear related to the ratios of the two heaviest
generations, which is expected if both come from a rotating mass matrix.

Finally, using the values of the parameters determined, one can calculate
also from (\ref{vapprox}) the 3 vectors ${\bf v}_\tau, {\bf v}_\mu$ and
${\bf v}_e$ for the charged leptons, as explained in the section above.
We notice in (\ref{vapprox}), however, that these vectors depend on the
angles $\phi_i$ which may not be small, not just on their differences.
These can be determined by solving equation (\ref{phieqn})
and using the result $\sin(\phi_s - \phi_c) \sim 0.18$.  
This gives 2 solutions, $\phi_s \sim 0.61, 0.35$, of which the fit in
\cite{phenodsm} correspond to the former.  With the values $\theta_t \sim
0.011$ obtained before and $\phi_s \sim 0.61$, one can then evaluate 
the corresponding angles for $\tau$ and $\mu$, and 
hence the vectors ${\bf v}_\tau, {\bf v}_\mu, {\bf v}_e$.  One obtains: 
\begin{eqnarray}
|\tau \rangle & = & (0.994, 0.070, 0.029), \nonumber\\
|\mu \rangle  & = & (-0.074, 0.757, 0.657), \nonumber\\
|e \rangle    & = & (0.0188, -0.657, 0.757),
\label{leptonvsa}
\end{eqnarray}
not far from the values quoted in (\ref{leptonvs}) from \cite{phenodsm},
and corresponding to the MNS lepton mixing matrix elements:
\begin{eqnarray}
U_{\mu 3} = \langle \mu|\nu_3 \rangle & = & 0.77, \nonumber\\ 
U_{e 3}   = \langle e  |\nu_3 \rangle & = & 0.07,
\label{Umu3e3tha}
\end{eqnarray}
which is seen to compare well with (\ref{Umu3e3th}) and (\ref{Umu3e3ex}) 
above.
 
Hence, one sees that all the previous results obtained numerically before
together with their agreement with experiment have now been confirmed by 
analytic considerations although in a rather crude approximation.

\vspace{1cm}

In summary, we conclude that a closer examination of the tenets of the
DSM scheme has removed a previous restriction on its predictions on 
neutrino oscillations, making them now consistent with the favoured
LMA solution for solar neutrinos.  It has also explained away the couple
of discrepancies noted before as limitations of the 1-loop calculation
so far performed, which is seen otherwise to be in full agreement with
experiment within its perceived range of validity.  Furthermore, all previous 
numerical results have now been confirmed by analytic considerations.


\begin{thebibliography}{99}

\bibitem{physcons} Chan Hong-Mo and Tsou Sheung Tsun, Phys. Rev. D57, 
   2507, (1998), hep-th/9701120.

\bibitem{dualgen} Chan Hong-Mo, talk given at the International Conference 
   on Fundamental Sciences, Mathematics and Theoretical Physics, 13-17 
   March 2000, Singapore, Int. Journ. Mod. Phys. A 16 (2001) 163-177, 
   hep-th/0007016.

\bibitem{genmixdsm} Chan Hong-Mo and Tsou Sheung Tsun, Lectures given at
   the 42nd Cracow School of Theoretical Physics, Zakopane, Poland,
   May-June, 2002, Acta Phys. Pol. B33 (2002) 4041.

\bibitem{ckm} Jos\'e Bordes, Chan Hong-Mo, Jacqueline Faridani, Jakov 
   Pfaudler, and Tsou Sheung Tsun,  Phys. Rev. D58, 013004, (1998), 
   hep-ph/9712276.

\bibitem{phenodsm} Jos\'e Bordes, Chan Hong-Mo and Tsou Sheung Tsun, 
   Eur. Phys. J. C. 10, 63 (1999), hep-ph/9901440.

\bibitem{databook} Review of Particle Physics, D.E. Groom et al., Eur.
   Phys. Journ. C15, 1, (2000).

\bibitem{SuperK} Superkamiokande data, see e.g. talk by T. Toshito at
   ICHEP'00, Osaka (2000).

\bibitem{Chooz}  CHOOZ collaboration, M. Apollonio et al., Phys.\ Lett.\ 
   B466, 415, (1999), hep-ex/9907037.

\bibitem{dualsymm} Chan Hong-Mo, Jacqueline Faridani, and Tsou Sheung Tsun, 
   Phys. Rev. D53, 7293 (1996), hep-th/9512173.

\bibitem{cevidsm} Jos\'e Bordes, Chan Hong-Mo and Tsou Sheung Tsun,
   hep-ph/0203124, to appear in Eur. Phys. J. C.

\bibitem{CKM}  N.\ Cabibbo, Phys.\ Rev.\ Lett.\ {\bf 10}, 531 (1963);
   M.\ Kobayashi and T.\ Maskawa, Prog.\ Teor.\ Phys.\ 49, 652 (1973).

\bibitem{MNS}  Z. Maki, M. Nakagawa and S. Sakata, Progr. Theor. Phys. 
   28, 870 (1962).

\bibitem{seesaw} M. Gell-Mann, P. Ramond, and S. SLansky in 
   {\it Supersymmetry}, edited by F. van Niuwenhuizen and D. Freeman
   (North Holland, Amsterdam, 1979); T. Tanagida, Prog. Theor. Phys.,
   B135, 66, (1978).

\bibitem{nuosc} Jos\'e Bordes, Chan Hong-Mo, Jakov Pfaudler, and 
   Tsou Sheung Tsun, Phys. Rev. D58, 053003, (1998), hep-ph/9802420.

\bibitem{SNO} Q.R. Ahmad et al. Phys. Rev. Lett. 87, 071307, (2001),
   nucl-ex/0106015.

\bibitem{MSW} L. Wolfenstein, Phys. Rev. D17, 2369, (1978); S.P. Mikheyev
   and A.Yu. Smirnov, Nuovo Cim. 9C, 17, (1986).

\bibitem{LMA} See e.g. J.N. Bahcall, M.C. Gonzalez-Garcia, and 
   C. Pe\~na-Garay, JHEP 0207 (2002) 054 (Aug. 6, 2002),
hep-ph/0204314.

\bibitem{kamland} K.\ Eguchi et al, hep-ex/0212021.

\bibitem{data03} Review of Particle Physics, K. Hagiwara et al.,
   Phys. Review D66, 010001 (2002).

\bibitem{2loop} Jos\'e Bordes, Chan Hong-Mo and Tsou Sheung Tsun, work
   in progress.

\bibitem{features} Jos\'e Bordes, Chan Hong-Mo, Jakov Pfaudler, and
   Tsou Sheung Tsun, Phys. Rev. D58, 053006, (1998), hep-ph/9802436.

\end{thebibliography}
\end{document}